\documentclass[a4paper,11pt]{article}
\usepackage{latexsym}
\usepackage{graphicx}
\usepackage{caption}
\usepackage{psfrag}
\usepackage{amsmath,amssymb,dsfont}
\usepackage{epsfig}
\newcommand{\be}{\begin{equation}}
\newcommand{\ee}{\end{equation}}
\newcommand{\bea}{\begin{eqnarray}}
\newcommand{\eea}{\end{eqnarray}}

\newcommand{\nn}{\nonumber}

\input epsf

\begin{document}

\begin{titlepage}

\begin{flushright}
\end{flushright}
\vspace*{1.5cm}
\begin{center}
{\Large \bf Pade Theory applied to the vacuum polarization \\ of a heavy quark}\\[3.0cm]

{\bf P. Masjuan, S. Peris} \\[1cm]

Grup de F{\'\i}sica Te{\`o}rica and IFAE\\ Universitat Aut{\`o}noma de Barcelona, 08193 Barcelona, Spain.\\[0.5cm]

\end{center}

\vspace*{1.0cm}

\begin{abstract}

The vacuum polarization of a quark, when considered in terms of the external momentum $q^2$, is a function of the Stieltjes type. Consequently, the mathematical theory of Pade Approximants assures that the full function, at any finite value of $q^2$ away from the physical cut, can be reconstructed from its low-energy power expansion around $q^2=0$. We illustrate this point by applying this theory to the vacuum polarization of a heavy quark and obtain the value of the constant  $K^{(2)}$  governing the threshold expansion at order $\mathcal{O}(\alpha_s^2)$.
\end{abstract}

\end{titlepage}

\section{Introduction}

One interesting object to study in connection with the physics of heavy quarks is the vacuum polarization function of two electromagnetic currents. This requires high order perturbative calculations which, because of the obvious need to keep a nonzero mass $m$ for the quark, become extremely difficult to perform. This is why, while the $\mathcal{O}(\alpha_s^0)$ and $\mathcal{O}(\alpha_s^1)$ contributions have been known for a long time \cite{Kallen}, state of the art calculations can only produce a result at $\mathcal{O}(\alpha_s^2)$ in the form of an expansion at low energies (i.e. $q^2=0$), at high energies (i.e.  $q^2=\infty$) or at threshold (i.e. $q^2=4 m^2$), but not for the complete function, which is still out of calculational reach. In this circumstances, it would of course be very interesting to be able to reconstruct this function by some kind of interpolation between the former three expansions.

After the work in refs. \cite{Broadhurst,ChetyrkinMethod,ChetyrkinH01}, it has become customary to attempt this
reconstruction of the vacuum polarization function with the help of  Pade Approximants. Since these
approximants are ratios of two polynomials in the variable $q^2$, they are very suitable for the
matching onto the low-energy expansion. This is so because this low-$q^2$ expansion is truly an
expansion in powers of $q^2$, as a consequence of the finite energy threshold\footnote{In
perturbation theory, this is true so long as purely gluonic intermediate states are not considered.
Beyond perturbation theory, the threshold occurs at $4 m^2_{\pi}$, where $m_{\pi}$ is the pion
mass. } starting at $4m^2$. However, it is clear that they cannot fully recover the nonanalytic
terms which appear, e.g., in the form of logarithms  of $q^2$  in the expansion at high energies
(or at threshold, where there is also a squared root behavior). Therefore, what is really done in
practice is to first subtract all these logarithmic pieces from the full function (impossible to
match exactly with a Pade) with the help of a \emph{guess} function with the appropriate
 threshold and high-energy behavior, and then apply Pades to the remaining regular
 expression\footnote{In fact, this is done after a conformal mapping whereby all the (cut)
 complex plane is mapped into a circle of unit radius.}. In this way, the authors of Ref. \cite{HoangMateu}
 were able to compute, e.g.,  the value of the constant  $K^{(2)}$ appearing in the $\mathcal{O}(\alpha_s^2)$
 expansion of the vacuum polarization  at threshold,  which has not yet been possible to obtain from a Feynman
 diagram calculation.

Although this result is very interesting, the construction is not unique.  As recognized in Ref.
\cite{HoangMateu}, some amount of educated guesswork is required in order to resolve the inherent
ambiguity in the procedure. For instance, a certain number of unphysical poles are encountered, and
some additional criteria have to be imposed in order to decide how to discard these poles. Since
the resulting ambiguity leads to a systematic error which needs to be quantified, this error is
then estimated by varying among several of the possible arbitrary choices in the construction.
Although all these choices are made judiciously and in a physically motivated manner, it is very
difficult to be confident of the error made in the result, which obviously has an impact on the
value extracted for the constant $K^{(2)}$.

In this short note we would like to point out that, regarding the vacuum polarization function,
one can do away with all the above ambiguities. The vacuum polarization function belongs to a
class of functions (the so-called Stieltjes functions) for which a well-known theorem assures
that the diagonal (and paradiagonal) Pade Approximants converge. The result of this theorem
together with the fantastic amount of information obtained on the Taylor expansion around
$q^2=0$, for which 30 terms are known \cite{Maier}, will allow us to predict a value for $K^{(2)}$.
As it turns out, our result is very close to that of Ref. \cite{HoangMateu}, although slightly smaller.

The Theory of Pade Approximants is sufficiently developed to make these approximants a systematic
mathematical tool. As we will see in this article, and has been exploited in Refs.
\cite{Broadhurst,ChetyrkinMethod,ChetyrkinH01, HoangMateu}, they can be very useful for higher order calculations
in perturbation theory. Furthermore, they can also be a very interesting conceptual tool for
large-$N_c$ QCD \cite{PerisPades}. In this case this is true even when the function is not
Stieltjes because large-$N_c$ QCD Green's functions are meromorphic and there are powerful theorems
in this case as well \cite{PerisMasjuan}. Finally, Pades can also be used for
analyzing the experimental data by fitting to a rational function rather than the more commonly
used polynomial fitting\cite{SC1}. They are also instrumental in discussing methods of
unitarization \cite{SC2}.

\section{Pades and the vacuum polarization}
Let us start by defining the vacuum polarization function $\Pi(q^2)$ through the correlator of two
electromagnetic currents $j^\mu(x)=\bar q(x)\gamma^\mu q(x)$,
\begin{align}
\label{pidef}
\left(g_{\mu\nu}q^2-q_\mu q_\nu\right)\, \Pi(q^2)
\, = \, \,
- \,i
\int\mathrm{d}x\, e^{iqx}\left\langle \,0\left|T\, j_\mu(x)j_\nu(0)\right|0\,
\right\rangle
\,,
\end{align}
where $q^\mu$ is the external four-momentum. As it is known, due to the optical theorem, the  $e^+e^-$ cross section is proportional to the imaginary part of $\Pi$. As a result, $\mbox{Im}\,\Pi$ is a positive definite function, i.e.
\begin{equation}
\label{positive}
\mbox{Im}\,\Pi(t+i \varepsilon)\geq 0 \ ,
\end{equation}
a property which will become crucial  in what follows.

In perturbation theory $\Pi(q^2)$ may be decomposed to ${\mathcal O}(\alpha_s^2)$ as
\begin{align}
\label{Pi}
\Pi(q^{2}) \, = \,&\,
\Pi^{(0)}(q^{2})
\, + \,\left(\frac{\alpha_{s}}{\pi}\right)\,
\Pi^{(1)}(q^{2})+\left(\frac{\alpha_{s}}{\pi}\right)^{2}\,
\Pi^{(2)}(q^{2})
\, + {\mathcal O}(\alpha_s^3)\, .
\end{align}
For definiteness, $\alpha_s$  denotes the strong coupling constant in the $\overline{\mathrm{MS}}$ scheme at the scale $\mu=m_{pole}$, but this is not important for the discussion which follows. Equation (\ref{Pi}) will be understood in the on-shell normalization scheme  where a subtraction at zero momentum has been made in such a way as to guarantee that $\Pi(0)=0$.

As it is well known, the vacuum polarization in Eq. (\ref{Pi}) satisfies a once subtracted dispersion relation, i.e.
\begin{equation}\label{disprel}
    \Pi(q^{2}) = q^2 \int_{0}^{\infty} \frac{dt}{t (t-q^2-i \varepsilon)}\ \frac{1}{\pi} \mbox{Im}\,\Pi(t+i \varepsilon)\ .
\end{equation}
Since all diagrams with intermediate gluon states are absent up to $\mathcal{O}(\alpha_s^2)$, the lower limit for the dispersive integral (\ref{disprel}) starts, in fact,  at a finite value given by the threshold for pair production, i.e. $4m^2$. This fact only carries over to higher orders in $\alpha_s$ provided these intermediate gluon states are neglected. From now on, we will restrict ourselves to the vacuum polarization in Eq. (\ref{Pi}) to $\mathcal{O}(\alpha_s^2)$, neglecting higher orders in $\alpha_s$.

In terms of the more convenient variable
\begin{equation}
\label{z}
z \, \equiv \, \frac{q^2}{4 m^2}\ ,
\end{equation}
one can rewrite Eq. (\ref{disprel}), after redefining $u=4 m^2/t$, as\footnote{We are simplifying the notation by replacing $ \Pi(4 m^2 z) \rightarrow  \Pi(z) $.}
\begin{equation}\label{disprelz}
    \Pi(z) = z \int_{0}^{1} \frac{d u}{1- u z-i \varepsilon}\ \frac{1}{\pi} \mbox{Im}\,\Pi\left(4 m^2 u^{-1}+i \varepsilon\right)\ .
\end{equation}
Recalling that a Stieltjes function is defined as\cite{Baker}\footnote{In Ref. \cite{Baker}, the variable is chosen to be $-z$  rather than $z$.}
\begin{equation}\label{Stieltjes}
    f(z)=\int_{0}^{1/R} \frac{d\phi(u)}{1-u z}
\end{equation}
where $\phi(u)$ is any \emph{nondecreasing} function, one sees that the identification

\begin{equation}\label{id}
    d\phi(u)=\frac{1}{\pi} \mbox{Im}\,\Pi\left(4 m^2 u^{-1}+i \varepsilon\right) \ du
\end{equation}
allows one to recognize that the integral in Eq. (\ref{disprelz}) defines the Stieltjes function $z^{-1}\Pi(z)$. As one can see, the positivity property Eq. (\ref{positive}) is crucial for the identification (\ref{id}) to be possible. The representation of the function $f(z)$ in Eq. (\ref{Stieltjes}) clearly shows a cut in the $z$ complex plane on the positive real axis for $R\leq z <\infty$. For the physical function $\Pi(z)$, this of course corresponds to the physical cut in momentum for $4m^2\leq q^2< \infty$, i.e. the physical case corresponds to $R=1$ in Eqs. (\ref{Stieltjes},\ref{id}). Furthermore, just like the function $f(z)$ in Eq. (\ref{Stieltjes}) has  a power series expansion convergent in the disk $|z|<R$, so does the function $ \Pi(z)$ in Eq. (\ref{disprelz}) have a power series expansion convergent in the disk $|z|<1$.

A Pade Approximant to a function $f(z)$, which will be denoted by  $P_N^M(z)$, is the ratio of two
polynomials of degree $M$ and $N$ (respectively)\footnote{Without loss of generality, the
denominator polynomial of degree $N$ is chosen to be unity at $z=0$.} such that its expansion in
powers of $z$ about the origin matches the expansion of the original function up to and including
the term of ${\mathcal O}(z^{M+N})$. When the original function $f(z)$ is Stieltjes with
a finite radius of convergence about the origin, $R$, it is a
well-known result in the theory of Pade Approximants that the sequence $P_N^{N+J}(z)$ (with $J\geq
-1$) converges to the original function, as $N\rightarrow\infty$, on any \emph{compact} set in the
complex plane, excluding the cut at $R\leq z <\infty$ \cite{Baker}. This excludes , in the physical case, the cut at $4m^2\leq q^2< \infty$ (recall that $R=1$). The position of the poles in the Pade
Approximant accumulate on the positive real axis starting at threshold, $q^2=4 m^2$, mimicking
the presence of the  physical cut in the original function. When Pades are applied to the vacuum
polarization, this means, in particular, that there can be no spurious pole outside of the positive
real axis in the $z$ plane and,
consequently, no room for ambiguities. Furthermore, the convergence of the approximation (and the
error) can be checked as a function of $N$, as we will see.

\section{Analysis}

In Eq. (\ref{Pi}), the full functions $\Pi^{(0,1)}(q^2)$ are known. They are given by the following expressions \cite{Kallen}:
 \begin{align}
\label{eq:Pi01}
\Pi^{(0)}(z) & \, = \,
\frac{3}{16\pi^{2}}\left[\frac{20}{9}+\frac{4}{3z}-\frac{4(1-z)(1+2z)}{3z}G(z)\right],
\nn \\[2mm]
\Pi^{(1)}(z) & \, = \,
\frac{3}{16\pi^{2}}\left[\frac{5}{6}+\frac{13}{6z}-\frac{(1-z)(3+2z)}{z}G(z)+
\frac{(1-z)(1-16z)}{6z}G^{\, 2}(z)\right. \nn  \\
&\qquad \qquad \qquad -\,\left.\frac{(1+2z)}{6z}\left(1+2z(1-z)\frac{d}{dz}\right)\frac{I(z)}{z}\right] \quad ,
\end{align}
where
\begin{align}
\label{eq:Gz}
I(z) & \, = \,
6\Big[\zeta_{3}+4\,\mbox{Li}_{3}(-u)+2\,\mbox{Li}_{3}(u)\Big]-
8\Big[2\,\mbox{Li}_{2}(-u)+\mbox{Li}_{2}(u)\Big]\ln u\nn \\
&\qquad \qquad \qquad \qquad  -2\Big[2\,\ln(1+u)+\ln(1-u)\Big]\ln^{2}u\,, 
\nn \\[2mm]
 G(z) & \, = \, \frac{2\, u\,\ln u}{u^{2}-1}\ ,
\quad
\mbox{with}\quad
 u \, \equiv \, \frac{\sqrt{1-1/z}-1}{\sqrt{1-1/z}+1}
\ .
\end{align}

 However, the situation with the function $\Pi^{(2)}(q^2)$ is different. In fact, $\Pi^{(2)}(q^2)$ is
 only partially known through its low-energy power series expansion around $q^2=0$, its high-energy
 expansion around $q^2=\infty$ and its threshold expansion around $q^2=4 m^2$, but the full function has
 not yet been computed. Unlike the latter two
 expansions, for which only a few terms are known, our knowledge of the expansion of $ \Pi^{(2)}(q^2)$ around $q^2=0$ is
 very impressive, after the work of Ref. \cite{Maier} where 30 terms of this expansion were computed.

Although the full vacuum polarization function $\Pi(q^2)$ is Stieltjes, there is no reason why all
the individual contributions $\Pi^{(0,1,2,...)}(q^2)$ should also have this property. Amusingly,
however, this happens to be true both for $\Pi^{(0)}(q^2)$ and $\Pi^{(1)}(q^2)$
\cite{Broadhurst}\footnote{The case of $\Pi^{(0)}(q^2)$ is trivial as it coincides with the full
vacuum polarization for $\alpha_s$=0.}. As we will now show, this is no longer the case for
$\Pi^{(2)}(q^2)$ because its power series expansion around $q^2=0$ does not satisfy certain
determinantal conditions which hold for a Stieltjes function.

Defining the power expansion around $z=0$ of the Stieltjes function $f(z)$ in Eq. (\ref{Stieltjes}) as
\begin{equation}\label{exp}
    f(z)= \sum_{n=0}^{\infty} f_n z^n\ ,
\end{equation}
the coefficients $f_n$ satisfy the following  determinantal conditions. Let $D(m,n)$ be the
determinant constructed with the Taylor  coefficients $f_n$
\begin{equation}\label{det}
    D(m,n)=\begin{vmatrix}
             f_m & f_{m+1} & \ldots & f_{m+n} \\
             f_{m+1} & f_{m+2} & \ldots & f_{m+n+1} \\
             \vdots & \vdots &  & \vdots \\
             f_{m+n} & f_{m+n+1} & \ldots & f_{m+2n} \\
           \end{vmatrix} \quad .
\end{equation}
 A Stieltjes function must satisfy $D(m,n)>0$, for all $m,n$ \cite{Baker}. However,
 using the  $f_n$ coefficients given in Ref. \cite{Maier} (in the on-shell scheme, with the number of light flavors $n_\ell=3$):
\begin{equation}\label{taylor}
    z^{-1} \Pi^{(2)}(z)\approx 0.631107 + 0.616294 \ z + 0.56596 \ z^2 + 0.520623 \ z^3+ \ldots \quad ,
\end{equation}
one can immediately see that, e.g.,
\begin{equation}\label{D11}
    D(0,1)=D(0,1)=\begin{vmatrix}
                             0.631107  &0.616294 \\
                              0.616294 &  0.56596 \\
                            \end{vmatrix}= -0.0226376  < 0\quad  .
\end{equation}
This proves that the individual function $\Pi^{(2)}(q^2)$ is, all by itself,  not a Stieltjes
function, even though the combination $\Pi(q^2)$ in (\ref{Pi}) is.  Therefore, we will now focus on
applying the Theory of Pade Approximants to the full combination $\Pi(q^2)$  in Eq. (\ref{Pi}) in
order to extract information on the individual term $\Pi^{(2)}(q^2)$.

Since, as it is obvious from Eq. (\ref{Pi}), the function $\Pi(q^2)$ depends
on the value of $\alpha_s$, any Pade Approximant to it will also depend on the
value of $\alpha_s$, i.e.  $P_N^{N+J}(z;\alpha_s )$. This means that it is possible
 to construct a rational approximation to the three functions $ \Pi^{(0,1,2)}(q^2)$
 from three different sequences of Pade Approximants to $\Pi(q^2)$ constructed at three
 arbitrary values of  $\alpha_s$, let us say $\alpha_s=0, \pm \beta$, with $\beta$
 sufficiently small so as to be able to neglect the terms of $\mathcal{O}(\alpha_s^3)$
  in Eq. (\ref{Pi}) . In this way one obtains
\begin{eqnarray}
\label{aprox}
  \Pi^{(0)}(z) &\approx & P_N^{N+J}(z;\alpha_s=0 ) \nn \\
  \Pi^{(1)}(z)  &\approx & \frac{\pi}{2 \beta}\left\{P_N^{N+J}(z;\alpha_s=\beta )-P_N^{N+J}(z;\alpha_s=-\beta ) \right\}\nn \\
   \Pi^{(2)}(z)  & \approx & \frac{\pi^2}{2 \beta^2}\left\{P_N^{N+J}(z;\alpha_s=\beta )+P_N^{N+J}(z;\alpha_s=-\beta )- 2  P_N^{N+J}(z;\alpha_s=0 )\right\}\, ,
\end{eqnarray}
  where $J\geq -1$ and $N\rightarrow \infty$. Since the value of $\beta$ chosen is arbitrary, the $N\rightarrow \infty$ limit should produce results which are independent of $\beta$,  due to the convergence of the Pade Approximants to $\Pi(q^2)$. Therefore, one should see that the three combinations (\ref{aprox}) are increasingly independent of $\beta$ as $N$ grows.\footnote{This independence of $\beta$ in the case of $ \Pi^{(0)}(q^2)$ is trivially true.} This is indeed what happens.

\begin{figure}
\renewcommand{\captionfont}{\small \it}
\renewcommand{\captionlabelfont}{\small \it}
\centering
\includegraphics[width=3in]{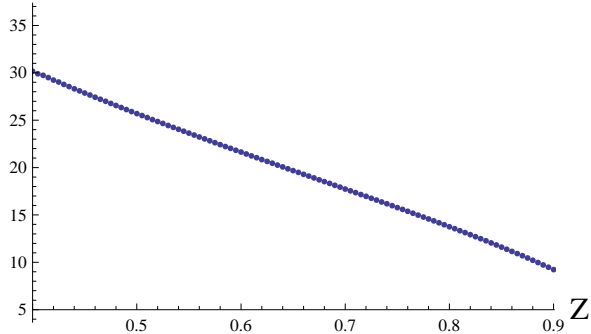}
\caption{Number of decimal places reproduced by the rational approximation in Eq. (\ref{aprox}) to the function $\Pi^{(1)}(z) $  as a function of $z$, in the interval $0.4\leq z\leq 0.9$. }\label{precision}
\end{figure}

Furthermore, since we know the exact function  $\Pi^{(1)}(z) $, we can compare it to the rational approximation on the right hand side of the second Eq. (\ref{aprox}) in order to test the approximation. Figure \ref{precision} shows the number of decimal places reproduced by this rational approximation in the interval $0.4 \leq z\leq 0.9$, when $N=14$ and $J=0$ (i.e. the diagonal Pade $P_{14}^{14}$),  for values of $\beta$ in the interval $0.1 \leq \beta \leq 1$. As one can see, the dependence on $\beta$ cannot be distinguished, and the accuracy reaches, e.g.,  $\sim 10$ decimal places at $z=0.9$.

The third Eq. (\ref{aprox}) yields the desired approximation to $\Pi^{(2)}(z) $. To be precise, it gives us a rational approximation to $\Pi^{(2)}(z) $ in any compact set of the $z$ complex plane, away from the cut $1\leq z< \infty$. Since the threshold expansion at $z\approx 1$ can be written as \cite{HoangMateu, Czarnecki}
\begin{eqnarray}\label{threshold}
    \Pi^{(2)}_{\mathrm{th.}}(z) &= &\frac{1.72257}{\sqrt{1-z}}
    +\left[0.34375-0.0208333\ n_{\ell}\right] \ \ln^{2}(1-z)\nn \\
    &+& \left[0.0116822\ n_{\ell} + 1.64058 \right]\ \ln(1-z) + K^{(2)}\nn \\
&&\!\!\!\!\!\!\!\!\!\!\!\!\!\!\!\!\!\!\!\!\!\!\!\!\!\!\!\!+ \left[-0.721213 - 0.0972614\ n_\ell +  3.05433 \ \ln(1-z)\right] \sqrt{1-z}
\, + \, {\mathcal O}(1-z)\ ,
\end{eqnarray}
in terms of an unknown constant $K^{(2)}$, our Pade Approximation (\ref{aprox}) may be used to determine this constant, as we will next discuss. In this threshold expansion we take $ n_{\ell}=3$ as the number of light flavors. Even though the numerical coefficients have been rounded off for simplicity, they may be extracted exactly from the results in Ref. \cite{Czarnecki}.

 Since Pades are not convergent on the physical cut, it is impossible to match the rational approximants (\ref{aprox}) to the threshold expansion (\ref{threshold}) as a function of $z$. This fact is obvious from the presence of logarithms and squared roots in Eq. (\ref{threshold}).  However, both approximations (\ref{aprox}) and (\ref{threshold}) are valid for values of $z$ at a finite distance from the cut and, in particular, in a certain window in the interval $0\leq z<1$. Within this window, a numerical matching of (\ref{aprox}) and (\ref{threshold}) is possible and will in fact  allow us to determine the unknown constant $K^{(2)}$.

 \begin{figure}
\renewcommand{\captionfont}{\small \it}
\renewcommand{\captionlabelfont}{\small \it}
\centering
\includegraphics[width=3in]{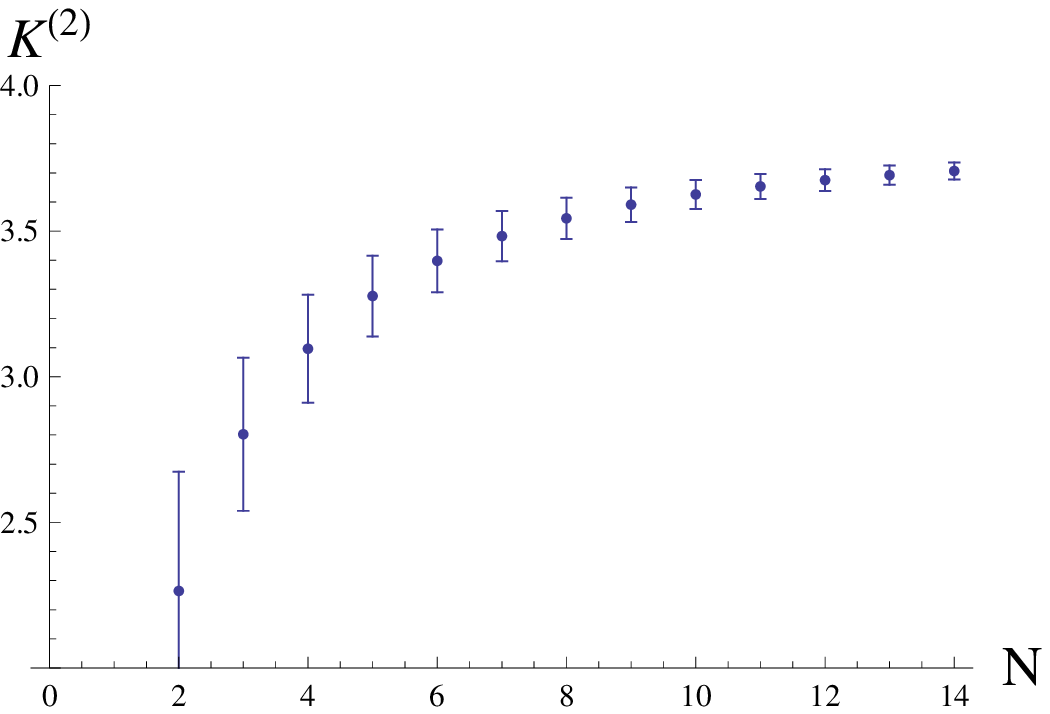}
\includegraphics[width=3in]{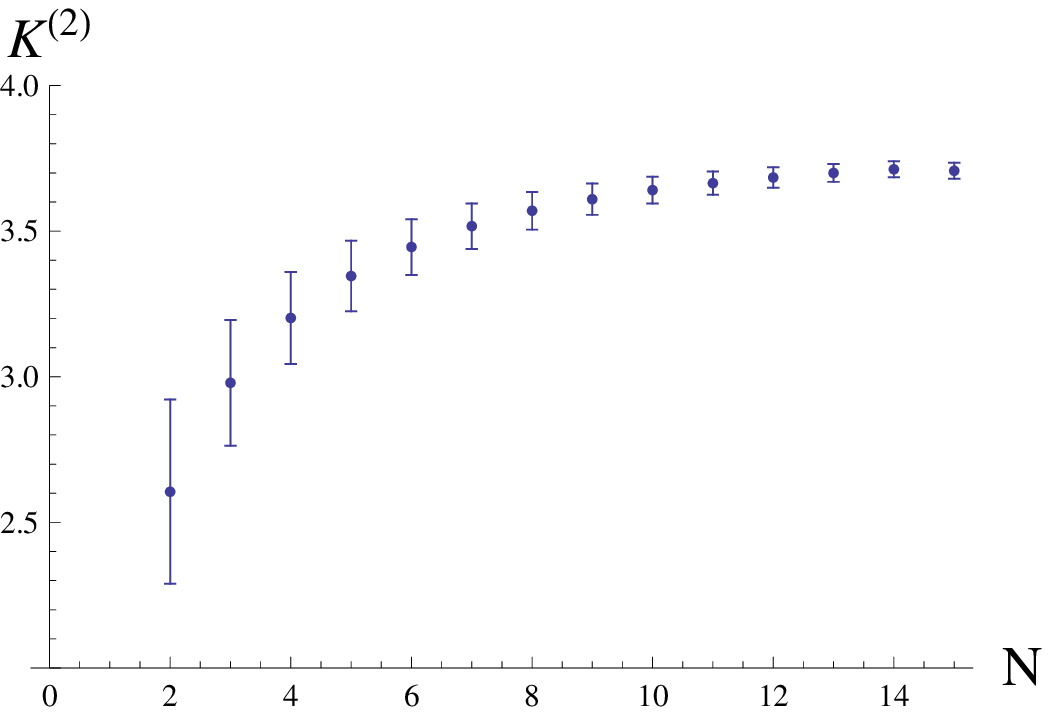}
\caption{Predicted value for $K^{(2)}$ from the sequence of diagonal, $P^{N}_{N}$ (left panel), and first paradiagonal, $P^{N-1}_{N}$ (right panel),  Pade Approximants. This figure corresponds to $\beta=0.5$.}\label{convergence}
\end{figure}

 In order to determine this window, we make the following observations. First, although the rational approximation (\ref{aprox}) is convergent as $N\rightarrow \infty$ in the interval $0\leq z< 1$, it is more accurate the closer one gets to $z=0$ in this interval, for a given value of $N$. On the other hand, the threshold expansion (\ref{threshold}) is more accurate the closer one gets to the branching point at $z=1$. From these two competing effects it is possible to determine an optimal window in $z$ by minimizing a combined error function. We will call this error function $\mathcal{E}(z)$.

 The function $\mathcal{E}(z)$ has to take into account the error from the Pades as well as the error from the threshold expansion. To estimate the error from the Pades, we consider the difference between two consecutive elements in the sequence, i.e.  $|P_{N}^{N+J}-P_{N-1}^{N-1+J} |$. As to the threshold expansion, we estimate its error as $|1-z|$, since the expression (\ref{threshold}) is accurate up to terms of $\mathcal{O}(1-z)$.  Therefore, in order to avoid possible accidental cancelations between the two errors, we define our combined error function as the following sum:
 \begin{eqnarray}\label{error}
   \!\!\!\! \mathcal{E}(z)&\!\!\!\!\!=&\!\!\!\!\!\left| \frac{\pi^2}{2 \beta^2}\Big\{P_N^{N+J}(z;\alpha_s=\beta )+P_N^{N+J}(z;\alpha_s=-\beta )- 2  P_N^{N+J}(z;\alpha_s=0 )\Big\}- \Big\{N\!\rightarrow \! N\!-\!1\Big\}\right|\nn\\
    &&\qquad +\quad |1-z|\ .
 \end{eqnarray}
Minimizing $\mathcal{E}(z)$ with respect
 to $z$ in the interval $0 \leq z< 1$, for every given values of $N$ and $\beta$, we may determine a
 value of $z$ at the minimum, namely $z^*$. This $z^*$ is then the one used to determine the constant $K^{(2)}$ as
 \begin{equation}\label{K}
   \!\!\! K^{(2)}\approx \frac{\pi^2}{2 \beta^2}\left\{P_N^{N+J}(z^*;\alpha_s=\beta )+P_N^{N+J}(z^*;\alpha_s=-\beta )- 2  P_N^{N+J}(z^*;\alpha_s=0 )\right\}- \widehat{\Pi}^{(2)}_{\mathrm{th.}}(z^*) \quad ,
 \end{equation}
 for the given $N$ and $\beta$. In Eq. (\ref{K}), $ \widehat{\Pi}^{(2)}_{\mathrm{th.}}(z^*) $ stands for the expression in Eq. (\ref{threshold}) without the constant $K^{(2)}$ and, of course, without the term $\mathcal{O}(1-z)$, evaluated at $z=z^*$. The knowledge of 30 terms from the
 low-energy expansion gives us enough information  to be able to construct up to
  the Pade $P_{14}^{14}$ from the diagonal sequence, and up to the Pade $P_{15}^{14}$
  from the first paradiagonal sequence. This corresponds to  $J=0$ and $J=-1$ in Eq. (\ref{K}). In all
  cases considered we have varied $\beta$ in the generous range $0\leq \beta \leq 1$, but our results
   are insensitive to this variation within errors, as expected.

\begin{figure}
\renewcommand{\captionfont}{\small \it}
\renewcommand{\captionlabelfont}{\small \it}
\centering
\includegraphics[width=3in]{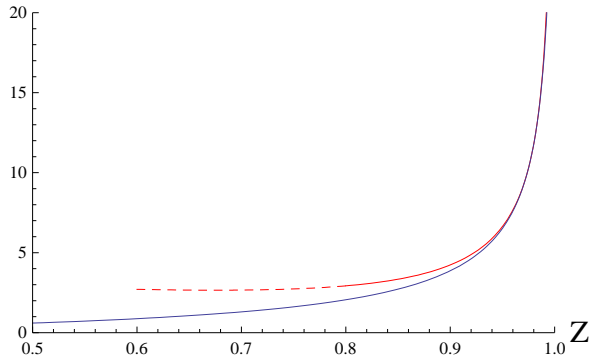}
\caption{Matching of the rational approximant in Eq. (\ref{aprox}) (solid blue line) to the
threshold expansion $\Pi^{(2)}_{\mathrm{th.}}(z)$ in (\ref{threshold}) (solid-dashed red line)  for
$N=14, J=0$ and the value of  $K^{(2)}$ in (\ref{result}). This figure shows the result for different values of $\beta$ in the range $0\leq \beta\leq1$, but the dependence on $\beta$ is so small
that cannot be discerned.}\label{matching}
\end{figure}

\section{Result}

 Our results for the constant $K^{(2)}$ are shown in Fig. \ref{convergence}. This figure shows
 the convergence of the diagonal sequence and the first paradiagonal sequence as a function
 of the order in the Pade. As one can see, we find a very nice convergence in both cases, with compatible results. Based on
this analysis, we obtain the following  value of  $K^{(2)}$:
\begin{equation}\label{result}
    K^{(2)}=3.71\pm 0.03\ .
\end{equation}
This result is very close to, although slightly smaller than, the value
 obtained in Ref. \cite{HoangMateu}, i.e. $K^{(2)}=3.81\pm 0.02$. This is the main result of this work.

 The error bars shown on Fig. \ref{convergence} have been  calculated as $\pm
 \mathcal{E}(z^*)$. Looking at the figure we see that the change in the value of  $K^{(2)}$ from
 one element of the sequence to the next is included in the errors shown, which we interpret as a
 sign that the estimate for the error we have made is rather accurate. Although we have taken symmetric errors
 for simplicity, it is also clear from the figure that the approach to the true value is made from
 below, so that a slightly more accurate determination could be achieved with the use of an asymmetric
 error. Apart from that, given the present knowledge of the expansions at low energy (\ref{taylor}) and at
 threshold (\ref{threshold}), we find it difficult to believe any error estimate which could significantly go below
 our figure in Eq. (\ref{result}). Of course, should more terms in either expansion be known, a
 rerun of  our analysis could immediately produce a more precise determination of  $K^{(2)}$.

Figure \ref{matching} shows the matching of the rational approximant (i.e. the right hand side of
the third of the Eqs. (\ref{aprox})) to the threshold expansion given by
$\Pi^{(2)}_{\mathrm{th.}}(z)$ in Eq. (\ref{threshold}), for $N=14$ and $J=0$,
 i.e. with the Pade $P_{14}^{14}$, and for the value of $K^{(2)}$ we have obtained. As one can see,
 this Pade is able to reproduce, with high accuracy, the threshold expansion behavior in
 a window $0.92 \lesssim z < 1$. At $z=1$ and above, the two lines in Fig. \ref{matching} will again diverge from
 each other, just as they do at low $z$. The  value of $z^*$ minimizing the error function $\mathcal{E}(z)$ in (\ref{error}) was found at $z^* \simeq 0.98$ in this particular case.

\begin{figure}
\renewcommand{\captionfont}{\small \it}
\renewcommand{\captionlabelfont}{\small \it}
\centering
\includegraphics[width=3in]{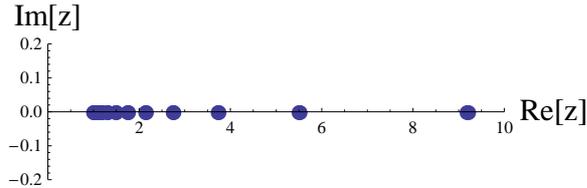}
\caption{Location of the poles in the pade $P_{14}^{14}$ in the complex plane. Notice the
accumulation of poles at $z\gtrsim 1$, simulating the physical cut. }\label{poles}
\end{figure}

For illustration, in Fig. \ref{poles} we show the position of the poles in the Pade $P_{14}^{14}$. As one can see, all the poles are sitting on the positive real axis above $z=1$, as it should be. Notice how they
accumulate in the region $z\gtrsim 1$. This is how Pades approximate the physical cut present
in the original function. As ensured from Pade Theory, this behavior was found in all the Pades
considered.

Finally, as a further test of our method, we have calculated the value of the constants $H^{(2)}_0$ and
$H^{(2)}_1$ which appear in the large-$z$ expansion of the function $\Pi^{(2)}(z)$ (we take
$n_f=n_{\ell}+1=4$ in the following expression):

\begin{eqnarray}
\label{Pihigh}
&&\Pi_{High-z}^{(2)}(z) \,= \,  \,
(0.034829 - 0.0021109\ n_f)\ln^{2}(-4z)
+(-0.050299 + 0.0029205\ n_f)\ln(-4z) \nn \\
& &+\quad  H^{(2)}_0
+ (0.18048 - 0.0063326\ n_f)\frac{\ln^{2}(-4z)}{z}
+ (-0.59843 + 0.027441\ n_f)\frac{\ln(-4z)}{z}\nn \\
& &+\quad  \frac{H^{(2)}_1}{z}
+\,{\mathcal O}\Big(z^{-3} \ln^3(-z)\Big)\ .
\end{eqnarray}

Using our method, we find $H_0^{(2)}=-0.582 \pm 0.008$. This result is to be compared to the true value $H^{(2)}_0=-0.5857$ \cite{ChetyrkinH01}. If we now input this exact value of $H_0^{(2)}$, by a rerun of the method, we may then determine the value of  $H^{(2)}_1$. In this way, we find  $H_1^{(2)}=-0.194 \pm 0.033$, which is to be compared to the exact value $H^{(2)}_1=-0.1872$ \cite{ChetyrkinH01}. Again, we find this agreement rather reassuring.

\vspace{1cm}
\textbf{Acknowledgements}

We thank A. Pineda and J.J. Sanz-Cillero for discussions and A. Hoang and V. Mateu for comments on the manuscript. This work has been supported by
CICYT-FEDER-FPA2008-01430, SGR2005-00916, the Spanish Consolider-Ingenio 2010
Program CPAN (CSD2007-00042)
and by the EU Contract No. MRTN-CT-2006-035482, ``FLAVIAnet''.


\begin{thebibliography}{99}

\bibitem{Kallen}
  A.~O.~G.~Kallen and A.~Sabry,
  Kong.\ Dan.\ Vid.\ Sel.\ Mat.\ Fys.\ Med.\  {\bf 29N17} (1955) 1.

\bibitem{Broadhurst}
  D.~J.~Broadhurst, J.~Fleischer and O.~V.~Tarasov,
  Z.\ Phys.\  C {\bf 60} (1993) 287
  [arXiv:hep-ph/9304303];
  J.~Fleischer and O.~V.~Tarasov,
  Z.\ Phys.\  C {\bf 64} (1994) 413
  [arXiv:hep-ph/9403230];
  P.~A.~Baikov and D.~J.~Broadhurst,
  arXiv:hep-ph/9504398.


\bibitem{ChetyrkinMethod}
  K.~G.~Chetyrkin, J.~H.~Kuhn and M.~Steinhauser,
  Phys.\ Lett.\  B {\bf 371} (1996) 93
  [arXiv:hep-ph/9511430].


\bibitem{ChetyrkinH01}
  K.~G.~Chetyrkin, J.~H.~Kuhn and M.~Steinhauser,
  Nucl.\ Phys.\  B {\bf 482} (1996) 213
  [arXiv:hep-ph/9606230].

\bibitem{HoangMateu}
  A.~H.~Hoang, V.~Mateu and S.~Mohammad Zebarjad,
  arXiv:0807.4173 [hep-ph].


\bibitem{Maier}
  A.~Maier, P.~Maierhofer and P.~Marquard,
  Nucl.\ Phys.\  B {\bf 797} (2008) 218
  [arXiv:0711.2636 [hep-ph]]; see, in particular, Table A.2.




\bibitem{PerisPades}
  S.~Peris,
  Phys.\ Rev.\  D {\bf 74} (2006) 054013
  [arXiv:hep-ph/0603190].

\bibitem{PerisMasjuan}
 P.~Masjuan and S.~Peris,
  JHEP {\bf 0705} (2007) 040
  [arXiv:0704.1247 [hep-ph]];
  Phys.\ Lett.\  B {\bf 663} (2008) 61
  [arXiv:0801.3558 [hep-ph]];
   P.~Masjuan,
  arXiv:0809.2704 [hep-ph].


\bibitem{SC1}
  P.~Masjuan, S.~Peris and J.~J.~Sanz-Cillero,
  Phys.\ Rev.\  D {\bf 78}, 074028 (2008)
  [arXiv:0807.4893 [hep-ph]].



\bibitem{SC2}
  P.~Masjuan, J.~J.~Sanz-Cillero and J.~Virto,
  arXiv:0805.3291 [hep-ph].

\bibitem{Baker}
  G.A.~Baker and P.~Graves-Morris,
  {\it Pad\'{e} Approximants, Encyclopedia of Mathematics and its
  Applications}, Cambridge Univ. Press. 1996, chapter 5;
 C.~Brezinski and J.~Van Inseghem,
 {\it Pad\'{e} Approximations, Handbook of Numerical Analysis},
 P.G. Ciarlet and J.L. Lions (editors), North Holland, vol. III;
 see also, e.g., C. Diaz-Mendoza, P. Gonzalez-Vera and R. Orive,
 Appl. Num. Math. \textbf{53} (2005) 39 and references therein.
  For a very pedagogical summary, see
C. Bender and S. Orszag, \textit{Advanced Mathematical Methods for Scientists and
Engineers I: asymptotic methods and perturbation theory}, Springer 1999, section 8.6



\bibitem{Czarnecki}
  A.~Czarnecki and K.~Melnikov,
  Phys.\ Rev.\ Lett.\  {\bf 80} (1998) 2531
  [arXiv:hep-ph/9712222].




\end{thebibliography}
\end{document}